\documentclass[prl,aps,twocolumn,superscriptaddress]{revtex4-1}
\usepackage{amssymb,amsfonts,amsmath,graphicx,color,times,mathtools}
 \pdfoutput=1
\usepackage{subfig}
\usepackage[normalem]{ulem}
\usepackage{hyperref}

{\catcode`\|=\active 
  \gdef\Braket#1{\begingroup
\mathcode`\|32768\let|\BraVert\left<{#1}\right>\endgroup}}
\def\BraVert{\egroup\,\mid\,\bgroup}

\newcommand{\be}{\begin{eqnarray}}
\newcommand{\ee}{\end{eqnarray}}
\newcommand{\ket}[1]{\left|#1\right\rangle}
\newcommand{\bra}[1]{\left\langle #1\right|}

\def\bra#1{\langle #1 |}
\def\ket#1{| #1 \rangle}

\DeclareMathOperator{\sgn}{sgn}

\definecolor{john}{rgb}{0.3,0.3,0.8}

\definecolor{alex}{rgb}{0,0.5,0.9}

\definecolor{tb}{rgb}{0.5,0.5,0.9}

\allowdisplaybreaks

\begin{document}

\title{Dynamical phase transitions, temporal orthogonality and the dynamics of observables in one dimensional ultra-cold quantum gases: from the continuum to the lattice}

\author{Thom\'{a}s Fogarty}
\email{thomas.fogarty@oist.jp}
\affiliation{Quantum Systems Unit, Okinawa Institute of Science and Technology Graduate University, Onna, Okinawa 904-0495, Japan}

\author{Ayaka Usui}
\email{ayaka.usui@oist.jp}
\affiliation{Quantum Systems Unit, Okinawa Institute of Science and Technology Graduate University, Onna, Okinawa 904-0495, Japan}

\author{Thomas Busch}
\email{thomas.busch@oist.jp}
\affiliation{Quantum Systems Unit, Okinawa Institute of Science and Technology Graduate University, Onna, Okinawa 904-0495, Japan}

\author{Alessandro Silva}
\email{asilva@sissa.it}
\affiliation{SISSA, Via Bonomea 265, I-34135 Trieste, Italy}

\author{John Goold}
\email{jgoold@ictp.it}
\affiliation{The Abdus Salam International Centre for Theoretical Physics (ICTP), Trieste, Italy}

\date{\today}

\begin{abstract}
We investigate the dynamics of the rate function and of local observables after a quench in models which exhibit phase transitions between a superfluid and an insulator in their ground states. Zeros of the return probability, corresponding to singularities of the rate functions, have been suggested to indicate the emergence of dynamical criticality and we address the question of whether such zeros can be tied to the dynamics of physically relevant observables and hence order parameters in the systems. For this we first numerically analyze the dynamics of a hard-core boson gas in a one-dimensional waveguide when a quenched lattice potential is commensurate with the particle density. Such a system can undergo a pinning transition to an insulating state and we find non-analytic behavior in the evolution of the rate function which is indicative of dynamical phase transitions. In addition, we perform simulations of the time dependence of the momentum distribution and compare the periodicity of this collapse and revival cycle to that of the non-analyticities in the rate function: the two are found to be closely related only for deep quenches. 
We then confirm this observation by analytic calculations on a closely related discrete model of hard-core bosons in the presence of a staggered potential and find expressions for the rate function for the quenches. 
By extraction of the zeros of the Loschmidt amplitude we uncover a non-equilibrium timescale for the emergence of non-analyticities and discuss its relationship with the dynamics of the experimentally relevant parity operator.
 \end{abstract}

\maketitle

\makeatletter

\section{Introduction} Recent experimental progress has reached a state where the dynamics of a complex and thermally isolated quantum system can be studied for unprecedentedly long evolution times. In particular, advances in the field of ultra-cold atoms have allowed for such a high degree of controllability  that, when combined with the absence of thermal phonons, studies of non-equilibrium coherent dynamics over timescales which are usually inaccessible in conventional condensed matter physics are possible \cite{bloch2008many,langen2015ultracold}. Not surprisingly, this has inspired a surge of theoretical interest and a growth of whole scientific communities which aim at the description of isolated, non-equilibrium, quantum systems~\cite{dziarmaga2010dynamics,polkovnikov2011colloquium,cazalilla2011one,eisert2015quantum,d2016quantum}. 

Pioneering early experiments in this direction included the observation of the non-equilibrium dynamics of a one dimensional Bose gas (a paradigmatic integrable model) \cite{kinoshita2006quantum}, which reopened foundational issues regarding thermalisation of observables in closed quantum systems \cite{polkovnikov2011colloquium,eisert2015quantum,d2016quantum}. Perhaps the earliest experiment in this field was conducted by Greiner {\it et al.} \cite{greiner2002collapse}, where a system was quenched across a superfluid to Mott-insulator transition and a coherent collapse and revival of the interference peaks in momentum space was observed in real time. This highly non-trivial non-equilibrium dynamics will be a central focus of this work and we aim at investigating its relationship to theoretical work which highlights the emergence of dynamical phase transitions (DPTs) in quenched dynamics. The idea of DPTs was first introduced by Heyl {\it et al.}, who studied the vacuum persistence amplitude (survival probability) for certain quenches in the paradigmatic transverse Ising model \cite{heyl2013dynamical}. Through a well known mapping with the boundary partition function \cite{leclair1995boundary,silva2008statistics} they noticed that the rate function for certain quenches exhibits non-analyticities whenever the wave function becomes orthogonal to the initial state. According to Heyl {\it et al.} this behavior therefore indentifies a dynamical phase transition. Since the original inception, DPTs have been studied in a wide range of models~\cite{karrasch2013dynamical,gong2013prethermalization,fagotti2013dynamical,pozsgay2013dynamical,andraschko2014dynamical,canovi2014first,heyl2014dynamical,hickey2014dynamical,kriel2014dynamical, vajna2014disentangling,heyl2015scaling,palmai2015edge, sharma2015quenches,abeling2016quantum,bhattacharya2016interconnections, heyl2016quenching,huang2016dynamical,maraga2016linear,sharma2016slow,vzunkovivc2016dynamical,zvyagin2016dynamical,yang2016,campbell2016,zunkovic2016dynamical,karrasch2017dynamical,obuchi2017complex} and while originally DPTs were believed to manifest when the system was quenched across an equilibrium phase transition, it is now known that they can occur even for quenches within the same phase~\cite{hickey2014dynamical,andraschko2014dynamical,vajna2014disentangling}. An exciting recent development is the observations of DPTs in experimental platforms such as ion trap architectures \cite{jurcevic2016direct} and cold atom arrays~\cite{flaschner2016observation}. 

Despite the range of models that have been investigated in relation to DPTs over the past years it is perhaps surprising that there have been little or no investigations of their manifestation in the original experiments which ignited the field, i.e.~the breathing dynamics across the superfluid to Mott insulator transition~\cite{greiner2002collapse} and dynamics in the Tonks-Girardeau gas~\cite{kinoshita2006quantum}. One central aim of this work is to fill that void. For this we first clarify the meaning of non-analyticities in the rate function proposed in \cite{heyl2013dynamical} and show then that, in general, the orthogonality of the time evolved state to the initial state is not related to the temporal behaviour of local observables. 
Our first system of choice for this is an important continuum model,  namely the Tonks-Girardeau gas \cite{girardeau1960relationship} undergoing a pinning transition to an insulator by application of a commensurate lattice potential. This effect was first theoretically predicted by B\"uchler {\it et al.} \cite{buchler2003commensurate} and later experimentally realized by Haller {\it et al.} \cite{haller2010pinning}. The dynamical quench problem was first studied by Lelas {\it et al.} in \cite{lelas2012pinning}. In our calculations we provide the first evidence of periodically appearing non-analyticities in the rate function for this process and explore the connection to the collapse/revival cycles in the dynamics of the momentum distribution. Both periodic cycles turn out to be connected only for deep quenches. 

We then confirm this observation by presenting an exactly solvable discrete model which contains the same physical phenomenology i.e.~hard-core bosons in a lattice at half filling with a staggered field. In this model analytic expressions can be found for the rate function and we compute the dynamics of the experimentally relevant parity operator and detail the connection with the rate function. 

In the following we will first briefly review the basic ideas relating to DPTs and particular the connection with dynamical restoration of symmetry. We then first present our results for the continuum model and follow this with an in-depth discussion of the lattice model. After this we conclude with an overall discussion of some of the issues raised.   

\section{Dynamical Phase Transitions}
 
 The DPTs defined by Heyl {\it et al.} \cite{heyl2013dynamical} are primarily centered around an object which is known as the Loschmidt amplitude
\begin{equation}
\label{amplitude}
	\mathcal{G}(t)=\bra\Psi e^{-iHt}\ket\Psi,
\end{equation}
and which has been exhaustingly studied under a number of guises in the past fifty years. 
This amplitude, following a Wick rotation $z=it$, can be thought of as a boundary partition function $\mathcal{Z}(z)=\bra\Psi e^{-z H}\ket\Psi$ for $z\in\mathcal{R}$ \cite{leclair1995boundary,silva2008statistics}. Exploiting this mapping, Heyl {\it et al.} noticed that, since the free energy density can be defined as $f(z)=-\lim_{L\rightarrow\infty}\frac{1}{L}\ln \mathcal{Z}(z)$ for a system of size $L$, 
the Fisher zeros in this boundary partition function (corresponding to singularities in $f(t)$) coalesce into lines which can cross the real axis. This leads to the emergence of critical times $t^{*}_n$ at which the so called rate function 
\begin{equation}
\label{rate}
	f(t)=-\frac{1}{L}\ln \mathcal{G}(t),
\end{equation}
displays non-analyticities.  
According to the definition of DPTs, these singularities identify points at which the time evolved state is orthogonal to the initial one and in the following we will examine this definition for analyzing the dynamics in systems which contain a superfluid-Mott insulator transition.

It is interesting to note that in the presence of symmetry breaking one can also modify the concept of dynamical criticality as the dynamical restoration of symmetry rather than orthogonality~\cite{heyl2014dynamical}. This can be seen by considering an
initial condition which breaks an $N_s$-fold symmetry of the Hamiltonian. Starting in $\ket{\Psi_0}$ and labeling the states obtained by repeated action of the symmetry operation as  $\{\bra{\Psi_j}\}$ ($j=1,..,N_s-1$), 
one can define the probability to remain in the ground-state manifold as
\be
	P(t)=\sum_{j=0}^{N_s-1}|\bra{\Psi_j} e^{-iHt} \ket{\Psi_0}|^2.
\ee
This quantity turns out to have singularities not in the presence of temporal orthogonality but  when the system crosses the boundary between two symmetry sectors. To demonstrate this let us consider for simplicity a twofold symmetry (like $Z_2$) and write according to eq.~\eqref{rate}
\be
	\bra{\Psi_j} e^{-iHt} \ket{\Psi_0}=e^{-Lf_{j}(t)},
\ee
where $f_0(t)$ and $f_1(t)$ correspond to the rate function in the two symmetry sectors. Let us now define the real valued rate function $l(t)=2\Re[f(t)]$. It is evident that at a certain time $t^*$ when the real parts of the rate function  coincide, ie $l_1(t)=l_0(t)$, the symmetry is dynamically restored, i.e.~there is equal probability to be in both symmetry sectors. At all other times one has  $l_1(t)>l_0(t)$ or $l_0(t)<l_1(t)$, which means that one of the two functions dominates $P(t)$ because the $L$ factor can be large in the exponentials. Therefore at  
the times $t^*$ cusp singularities appear in $P(t)$
and a correspondence between DPTs and standard symmetry breaking in the steady state can be established~\cite{zunkovic2016dynamical}. 

It is therefore clear that great care must be taken when interpreting non-analyticities in the rate function 
as points of dynamical criticality. Strictly speaking such non-analytic points are times when the evolving state after the quench becomes orthogonal to the initial state. Since, in general, this has nothing to do with the restoration of a symmetry one would not expect the global orthogonality to be reflected in the dynamics of experimentally relevant observables. However, as pointed out in \cite{heyl2013dynamical}, there is a case when they can be interpreted to be the same: if  the initial state is a Schr\"{o}dinger cat state of the form
\be
\label{cat}
	\ket{0}=\frac{1}{\sqrt{N_s}}\sum_{j=0}^{N_s-1}\;\ket{\Psi_j},
\ee 
i.e.~a linear superposition of symmetry related ground-states of the initial Hamiltonian.  Defining 
the generic rate functions via
\be
	\bra{\Psi_j} e^{-iHt} \ket{\Psi_k}=e^{-L f_{jk}(t)},
\ee
we get the Loschmidt amplitude
\be
	{\cal G}(t)=\bra{0} e^{-iHt} \ket{0}=\frac{1}{N_s}\sum_{j,k} e^{-Lf_{jk}(t)}.
\ee
Since in the thermodynamic limit this expression is dominated by the rate function with the smallest real part 
we have that 
\be
\lim_{L\rightarrow+\infty} |{\cal G}(t)|^2=P(t),
\ee
i.e. the return probability calculated on a state like Eq.\eqref{cat} is equivalent to the probability to stay in the ground state manifold in the thermodynamic limit. Since the \emph{Fisher zeroes} are singularities of the rate functions $f_{jl}(t)$, cusps in $P(t)$ emerge when two rate functions have the same real part. 

The two objects therefore generally give different information about the state of the system and the question is whether this information can be extracted from local measurements or not. Indeed, $P(t)$ can be shown to be connected to local symmetries of the Hamiltonian, since such symmetries are characterized by having local operators as generators. Furthermore, since the order parameter is an object which is in general \emph{not} invariant under such local operations, 
the cusps in $P(t)$ are naturally connected to zeroes of the order parameter since they indicate symmetry restoration. In turn singularities in ${\cal G}(t)$ and hence $f(t)$ (or equivalently $l(t)$) indicate orthogonality. Since the ground states of Hamiltonians across a symmetry breaking phase boundary are orthogonal (in the thermodynamic limit), it is  interesting to ask whether a connection between such singularities and the dynamics of local observables is present also in this case (see \cite{Jad1,Jad2} for a related study of criticality in systems with long range interactions). This is what we will investigate below in the first of the two models where we focus on the emergence of non analyticities in the rate function $f(t)$ in a highly experimentally relevant continuous model and explore their emergence with the dynamics of a measurable observable. 
 
\section{Temporal orthogonality in the Tonks-Girardeau gas }
The first model we consider describes a one-dimensional system of $N$ bosons confined in an external trapping potential. The Hamiltonian can be written as 
\begin{equation}
\label{eq:tg}
H=\sum_{j=1}^{N}\left(-\frac{\hbar^2}{2m}\nabla_j^2+V_{b}(x_j) + V(x_j) \right)+ g_{1\text{D}}\sum_{j>l}\delta(x_j-x_l),
\end{equation}
where $g_{1\text{D}}$ is a parameter characterizing the sign and magnitude of the interaction and $V_{b}(x)$ is a box potential of length $L$ with infinitely high walls. Let us assume an optical lattice potential of depth $V_0$ is applied in addition to the already existing trapping potential and is described by $V(x)=V_0\cos^2(k_0 x)$ where the wavevector is given by $k_0=M\pi/L$ and $M$ is the number of wells in the lattice. When the strength of the lattice is much larger than the recoil energy, $V_0\gg E_{r}=(\hbar k_0)^2/(2m)$, the model above can be mapped onto the celebrated Bose-Hubbard model, which has a transition between a superfluid and insulating state~\cite{bloch2008many}. In the limit when $V_0\ll E_{r}$, the Bose-Hubbard model is no longer applicable but interestingly it was shown in \cite{buchler2003commensurate} that at low energies the model can be mapped on to the Sine-Gordon model and a phase transition between a superfluid and insulating state remains when the applied lattice is commensurate with the particle density. The transition was observed experimentally by Haller {\it et al.} in 2010~\cite{haller2010pinning}. 

\begin{figure}[t!]
\includegraphics[width=0.99\columnwidth]{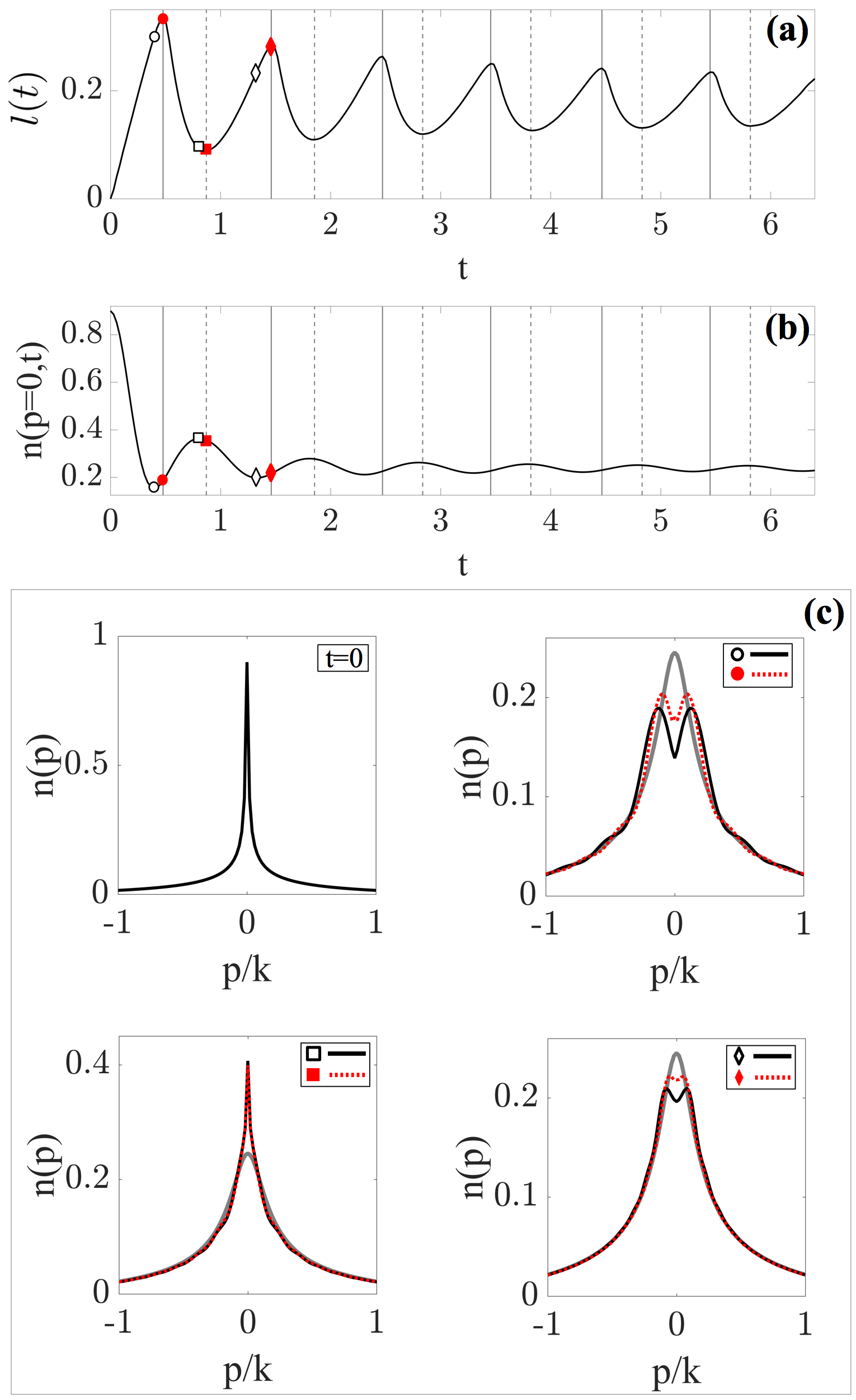}\\
\caption{Dynamics of a quench from the conducting ($V_i=0$) to the insulating phase ($V_f=E_R$) for a system of $N=100$ particles. (a) The rate function and (b) the height of the momentum peak as a function of time which is scaled with respect to $T_R$. The solid vertical lines indicate the times that the non-analyticities appear in $l(t)$, while the dashed vertical lines indicate the minima of $l(t)$. For the times marked by symbols in (a-b) the momentum distribution is plotted in (c). The black solid lines are for times corresponding to the extrema of the momentum peak, while the red dotted lines are for times corresponding to the extrema of $l(t)$. The grey solid line is the instantaneous momentum distribution of the insulating phase.}
\label{figON}
\end{figure}

In this work we will consider the hard-core limit of the system, $g_{1\text{D}}\rightarrow\infty$, where a pinning transition will occur for any infinitesimal lattice strength. In this limit the system is known as the Tonks-Girardeau gas, which allows for an exact solution due to the existence of the Fermi-Bose mapping theorem~\cite{girardeau1960relationship}. The essence of the mapping is that the interaction term in Eq.~\eqref{eq:tg} can be dealt with by imposing the following constraint on the many-body wave-function: $\Psi(x_1,x_2,\dots,x_{N})=0$ if $|x_{i}-x_{j}|=0$ for $i\neq j$ and $1\le i < j \le N$. The system can then be mapped to free fermions subject to appropriate symmetry: $\Psi(x_1,x_2,\dots,x_{N})=\prod_{1\le i < j \le N}\sgn(x_{i}-x_{j})\Psi_{F}(x_{1},x_{2},\dots,x_{N})$ where $\Psi_{F}=\frac{1}{\sqrt{N!}}\det^{N}_{n,j=1}[\psi_{n}(x_{j})]$ is a Slater determinant of single particle states.

This mapping theorem also holds time dependently and offers a convenient way to numerically calculate the real valued rate function $l(t)=2\Re[f(t)]$ from time evolving the single particle states in the quenched Hamiltonian $H_f$ as  
\begin{align}
l(t)&= -\frac{1}{L}\ln \left[|\langle \Psi_{0}|e^{-iH_ft} |\Psi_{0}\rangle|^{2}\right],\\
&=-\frac{1}{L}\ln\left[\det|A_{mn}(t)|^2\right],
\end{align}
where the $A_{mn}(t)=\int \psi^{*}_{m}(x,0)\psi_{n}(x,t)dx$ are the matrix elements of the overlaps between the pre- and post-quench single particle states. This allows for a straightforward and numerically exact approach to the computation of the rate function. 

The  figure of merit we will consider is the time dependent momentum distribution $n(p,t)$ which is routinely measured in cold atom setups. It is defined as the Fourier transform of the reduced single particle density matrix (RSPDM)
\begin{equation}
n(p,t)=\frac{1}{2\pi}\int dx dx' e^{ip(x-x')}\rho(x,x',t)
\end{equation} 
where the time dependent RSPDM is $\rho(x,x',t)=N\int dx_2\cdots dx_N \Psi^{*}(x',x_2,\dots,x_N,t)\Psi(x,x_2,\dots,x_N,t)
$, which is evaluated numerically using the technique developed in \cite{pezer2007momentum}. 

In the following we will study three types of quenches: switching the lattice on, switching  the lattice off and changing the sign of the lattice potential. 
If the lattice potential is commensurate with the particle number, $M=N$, then switching on the lattice potential from an initial depth $V_{i}=0$ to a final depth $V_{f}>0$ allows one to observe temporal orthogonality occurring in a quench from a conducting to an insulating phase. The rate function for this quench is shown in Fig.~\ref{figON}(a) and non-analytic peaks can be seen to occur at times $t/2+\alpha$ (where $\alpha$ is an integer) with a periodicity of $T_R=4\pi/V_f$. In panel (b) the value of the momentum distribution at $p=0$ is shown and for specific times the full momentum distribution is plotted in panel (c). The momentum distribution is initially sharply peaked at $p=0$, which is characteristic for a Tonks-Girardeau gas trapped in an infinite well and which reflects the expected partial first order coherence due to the order present in the RSPDM . After the quench the sharp peak vanishes as the momentum distribution broadens, signaling the transition to the insulating phase. The magnitude of the zero momentum component therefore oscillates as the system moves between insulating and conducting phases, with the first minimum occurring at a time which is slightly earlier than the emergence of the non-analytic peak in $l(t)$. For later times, this mismatch becomes more pronounced and the simulation clearly demonstrates
that the timescale for non-analyticities in the rate function quantifying orthogonality and that for the collapse/revival cycles in the momentum distribution are close but not the same. 
However,  the stronger the quench ($V_f >  E_R$), the more the two tend to coincide and we will explore this in more detail later when 
discussing the discrete model.

\begin{figure}
\includegraphics[width=0.99\columnwidth]{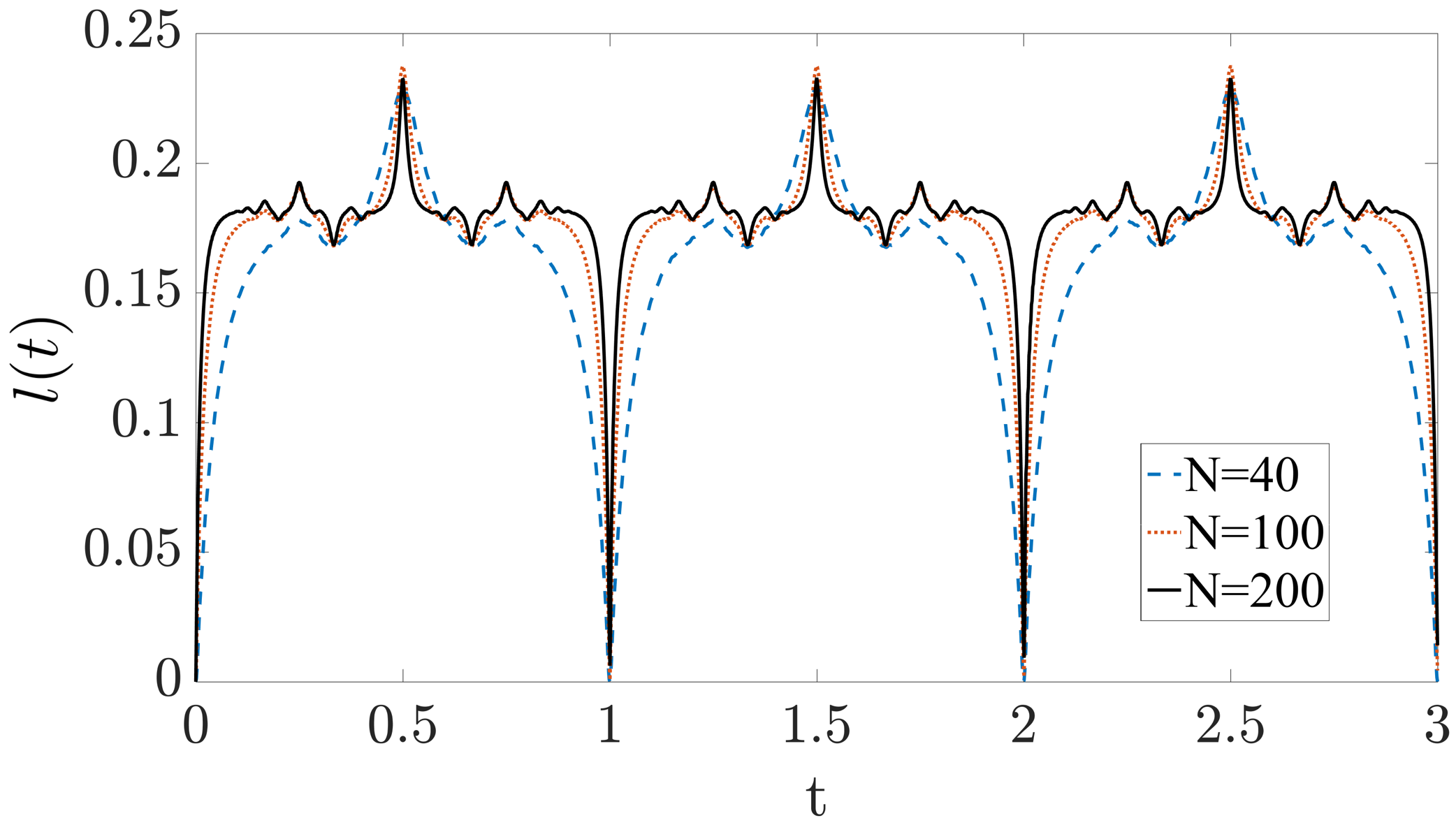}\\
\caption{Dynamics of a quench from insulator ($V_i=E_R$) to superfluid ($V_f=0$) for several systems with different particle number, $N$. Note that the time axis is rescaled by the revival time in the box, $N\pi/(2 E_r)$, which has the implication that the non-analyticities will not be observed in the thermodynamic limit.}
\label{figcradle}
\end{figure}

Let us now turn to the quench from insulator to superfluid, i.e.~from $V_{i}>0$ to $V_{f}=0$. The behaviour of the rate function is shown in Fig.~\ref{figcradle} for different system sizes on a time axis that is rescaled by $N\pi/(2 E_r)$. While one can observe a revival effect where at half the scaling time there is a type of transient criticality signalled by an apparent non-analyticity in $l(t)$ at times $t=\alpha+1/2$ ($\alpha$ an interger), these non-analyticities do not signal the existence of DPTs, but rather are a result of the propagation of density waves from the box edges which then interfere at the box center. This is precisely the dynamical de-pinning effect that was studied by Cartarius {\it et al} recently in the same model \cite{cartarius2015dynamical}. Therefore, this non-analyticity is the result of a finite size effect and does not exist in the thermodynamic limit. Instead the system undergoes a crossover from the insulating to the superfluid phase. 
This suggests that DPTs do not occur during dynamical de-pinning and we will explore this further in the discrete model in the next section. 

\begin{figure}
\includegraphics[width=0.99\columnwidth]{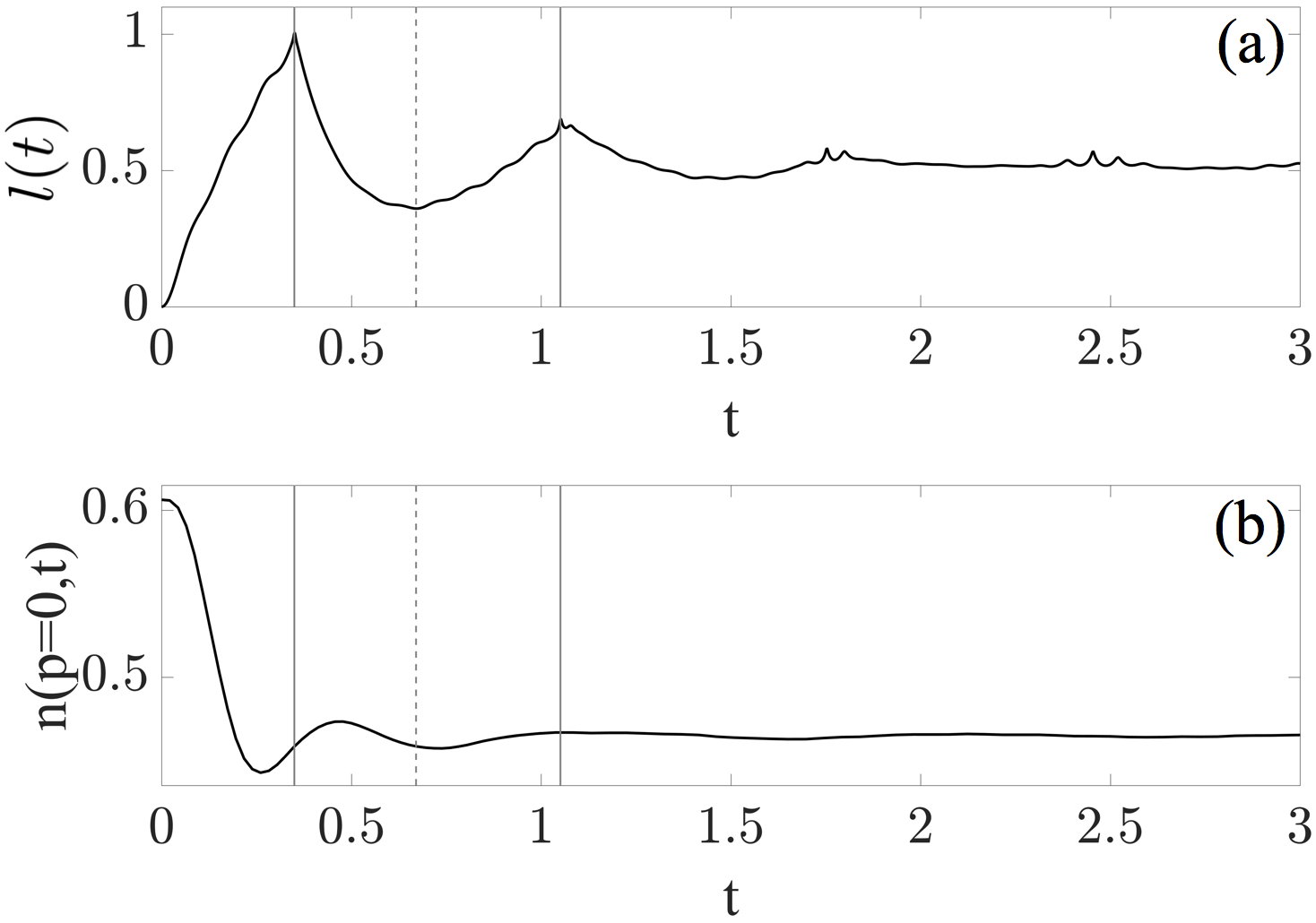}\\
\caption{(a) The rate function and (b) the height of the momentum peak for $N=100$ particles after a quench from the insulating phase ($V_i=2E_R$) to the insulating phase ($V_f=-2E_R$). The solid vertical lines indicate the times at which the non-analyticities appear in $l(t)$, while the dashed vertical line indicates its minima.}
\label{figVV2}
\end{figure}

Finally, we display in Fig.~\ref{figVV2} the dynamics of the rate function and the momentum distribution for quenches within the insulating phase, for $V_i=V$ to $V_f=-V$, which allows us to observe the post-quench dynamics on a timescale which is not governed by the lattice depth. Here the oscillations of both the rate function and momentum peak decay quickly, whilst it is clear that there is no simple relationship between non-analyticities which emerge in the rate function and any features in the behavior of the momentum distribution. Let us attempt to understand in detail this phenomenology by studying a closely related exactly solvable model.

\section{Tight binding model} We consider a system of $N$ hard core-bosons in a staggered onsite potential described by the Hamiltonian 
\begin{equation}
\label{britbastard}
H=J\sum_{j=1}^N\;(b^{\dagger}_j b_{j+1}+ h.c.) +\sum_{j=1}^N V(-1)^j b^{\dagger}_j b_j,
\end{equation}
where $b_j$ are hard core bosons, $J$ is the tunneling strength and $V$ is the strength of the onsite potential. This model has the distinct advantage over the previous continuous model in that it is analytically solvable while retaining all the essential physics. The procedure is well known \cite{lieb1961two}, using the Jordan-Wigner transformation $b^{\dagger}_j=e^{i\pi\sum_{l<j}a^{\dagger}_l a_l}a^{\dagger}_j$ and using Fourier transformed variables, $a_j=\frac{1}{\sqrt{N}}\sum_ke^{-ikj}a_k$ the Hamiltonian can be written as
\begin{eqnarray}
H=\sum_{|k|<\pi/2}\Psi^{\dagger}_k\;\hat{H}_k\;\Psi_k,
\end{eqnarray}
where $\Psi_k=(a_k,a_{k+\pi})^T$ and $\hat{H}_k=2J\cos(k)\mathbf{\sigma}^z+V\mathbf{\sigma}^y$,
where $\mathbf{\sigma}$ are the Pauli matrices. Notice that $k=\pi(2n+1)/N$, with $n=0,\dots, N/4-1$.
The Hamiltonian can be diagonalized in terms of the
new variables
$\Gamma_k=e^{i\theta_k\sigma^y}\Psi_k$,
where $\tan(2\theta_k)=V/(2\cos(k))$, and the resulting spectrum is characterized by a dispersion
$\epsilon_k=\sqrt{(2J\cos(k))^2+V^2}$. For our purposes we will  work at half filling where the spectrum is always gapped unless $V=0$, in which case the gap at $k =\pm \pi/2$ closes. Hence for $V \neq 0$ we have an insulating charge density wave phase while for $V=0$ it is a "superfluid". In what follows we will consider three different types of quenched dynamics as we did in the previous section: quenches from the superfluid to insulator, quenches from the insulator to superfluid and then quenches within the superfluid phase. We note that the same model can also be solved in the presence of an external flux~\cite{rossini2014quantum}.  

\begin{figure}[ht]
\centering
 \includegraphics[width=8cm]{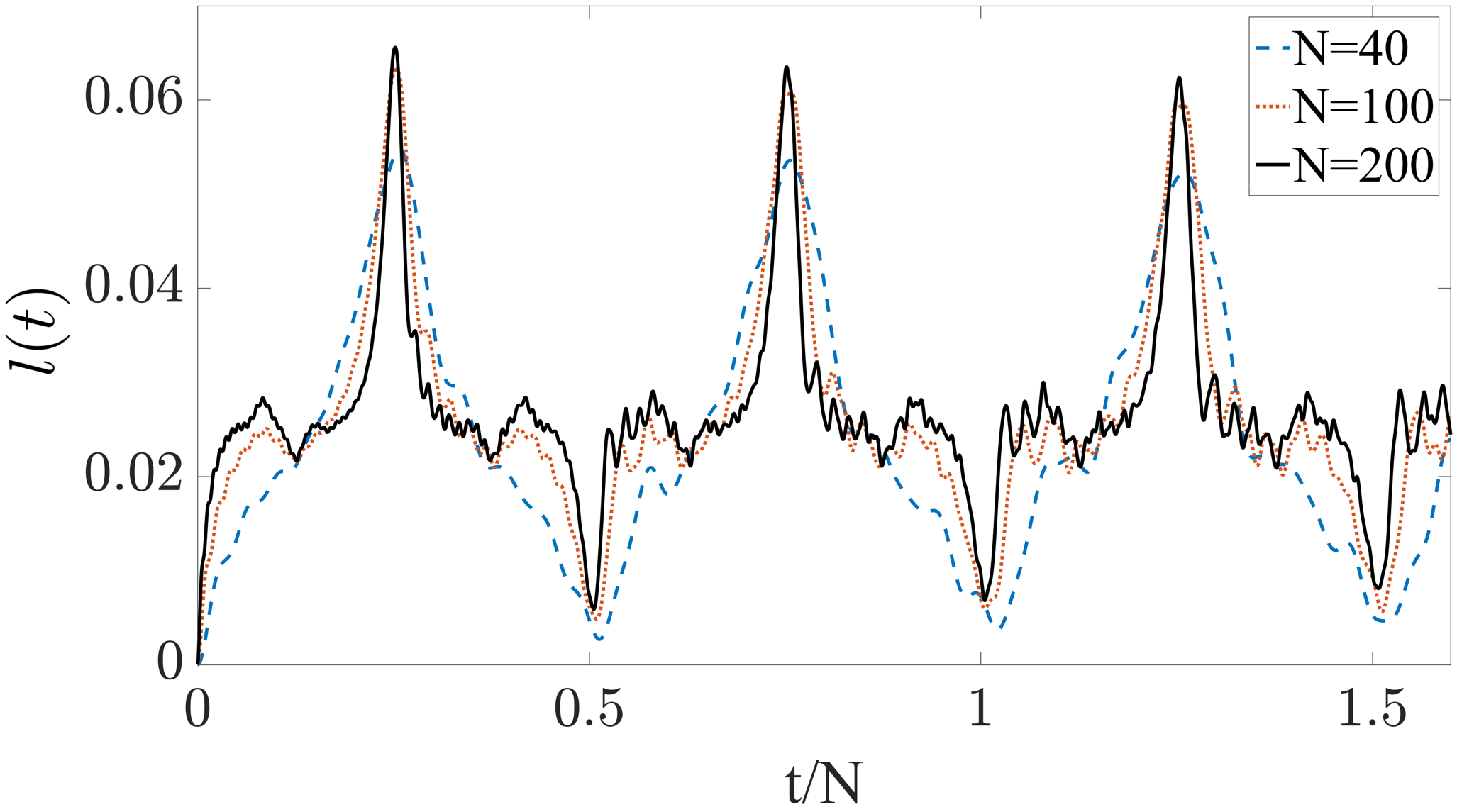}
\caption{$l(t)$ vs. $t/N$ for a quench from the insulator ($V_i=0.3$) to the superfluid ($V_f=0$) for different system sizes. }
\label{Figura1}
\end{figure}

Fixing the tunneling strength $J=1$ and considering a general quench from $V_i$ to $V_f$, the Loschmidt amplitude can be computed
using the Bogoliubov rotation connecting the old to the new quasiparticles 
$\Gamma_k(V_f)=\exp[i\Delta\theta_k\sigma_y]\;\Gamma_k(V_i)$ where $\Delta\theta_k=\theta_k(V_f)-\theta_k(V_i)$.
Representing the ground state $\ket{0}_{V_i}$ relative to $V_i$ as a squeezed state in terms of the $\Gamma_k(V_f)=|\gamma_+(k)\; \gamma_-^{\dagger}(k)|^T$ 
\begin{eqnarray}
\ket{0}_{V_i}=\frac{1}{{\cal N}} \prod_{|k|<\pi/2}\!\!(1+\tan(\Delta \theta_k) \gamma^{\dagger}_+(k)\gamma^{\dagger}_-(k)) \ket{0}_{V_f},
\end{eqnarray}
and computing the time evolution one finally obtains
\begin{eqnarray}
{\cal G}(z)=\!\!\!\!\prod_{|k|<\pi/2}\!\!\left\{ \frac{1+\tan(\Delta \theta_k)^2e^{2i\epsilon_k(V_f)z}}{1+\tan(\Delta\theta_k)^2}  \right\}.
\end{eqnarray}
Recalling that the Fisher zeroes are the roots of this complex valued function, one can solve them for ${\cal G}(z_{k})=0$ and find the expression
\begin{eqnarray}
z_k=\frac{(2n+1)\pi}{2\epsilon_k}+\frac{i}{\epsilon_k}\log(\tan(\Delta\theta_k)).
\end{eqnarray}
For quenches towards the insulating phase ($V_f>0$) it is evident that the Fisher zeros hit the real axis, hence corresponding to zeros of the Loschmidt amplitude (singularities of $l(t)$) whenever $\theta_k(V_f)-\theta_k(V_i)=\pi/4$. This corresponds 
to $\tan(2(\Delta\theta_k) =(2\cos(k)(V_f-V_i))/(4\cos(k)^2+V_f\;V_i)\rightarrow \infty$, which for
 $V_f>0$ and $V_i \in [-4/V_f,0]$ implies that $z_k=0$ for
\begin{eqnarray}
k^*=\arccos{\left[ \sqrt{-\frac{V_f\;V_i}{4}}\right]}.
\end{eqnarray}
A singularity at these momenta corresponds to a singularity in the rate function with a period
\begin{eqnarray}\label{period1}
T_R=\frac{\pi}{\sqrt{V_f(V_f-V_i)}}.
\end{eqnarray}
For quenches towards the superfluid phase Fisher zeroes have always a finite imaginary part implying the absence of 
singularities in $f(t)$ and therefore no DPTs are observed. However, keeping the system size finite and 
rescaling the time by it, one can observe a nice collapse and revival picture (see Fig.(\ref{Figura1})), as we previously discussed in the continuous model in Fig.~\ref{figVV2}(a).

\section{Orthogonality and observables}
As  discussed above, singularities in the rate function signal zeros at times when the time evolved state becomes orthogonal to the initial one. We now gain further understanding of why this occurrence is related to the time evolution of physical observables only for deep quenches. Notice that according to the calculation performed above,
the overlap between the different ground states of the Hamiltonian Eq.~\eqref{britbastard} at different strengths of the staggered potential, ${V_i}$ and ${V_f}$, is given by
\be
|\bra{\Psi_{V_f}} \Psi_{V_i}\rangle|^2=\exp\left[ -\frac{N}{2\pi}\int^{\pi/2}_{-\pi/2}dk\;\log[1+\tan(\Delta\theta_k)^2]   \right].\nonumber\\
\ee 
Therefore different ground states turn out to be \emph{orthogonal} in the thermodynamic limit, with the overlap vanishing exponentially in the system size. This is suggestive, since if upon quenching $V$ say from the superfluid $V_i=0$ to the insulator $V_f\neq 0$ the system dynamics would result in consecutive collapses and revivals of the superfluid into the insulator, one could expect the system to attain orthogonality with the initial state at the farthest point from the superfluid, i.e. when the collapse into the insulator is complete. This intuition would be correct if the system would be able to dissipate the work done on it by the quench procedure. In the the present case of unitary dynamics, however, the fact that the system remains  in a superposition of excited states of the post-quench Hamiltonian, makes the identification of the phenomena problematic. In other words it is only in the thermodynamic limit that ground states with different parameters are orthogonal. Hence only in that limit one could expect that, if the system was indeed able to collapse and turn back from one state to the other, one would get orthogonality when the superfluid fully collapses into a Mott insulator. An exception are deep quenches as we will now show.
\begin{figure}[ht]
\centering
 \includegraphics[width=8cm]{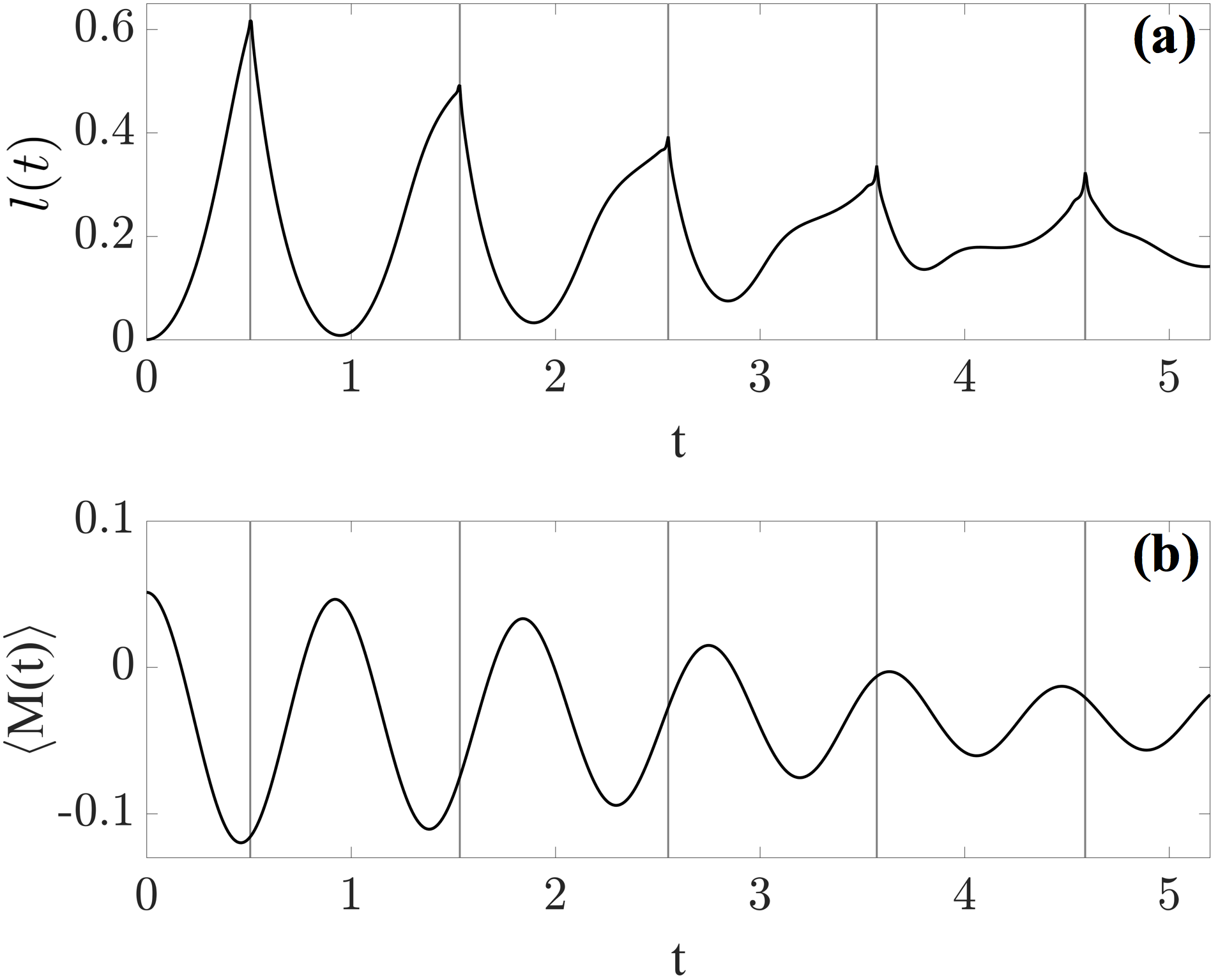}
\caption{(a) $l(t)$ and (b) $\langle M(t) \rangle$ vs. $t$ for $N=100$ and a quench from  $V_i=-1/6$ to $V_f=3$. The correspondence between minima of $\langle M(t) \rangle$ and cusps of $l(t)$ is not present in this case.
}
\label{Figure3}
\end{figure}

\begin{figure}[ht]
\centering
 \includegraphics[width=8cm]{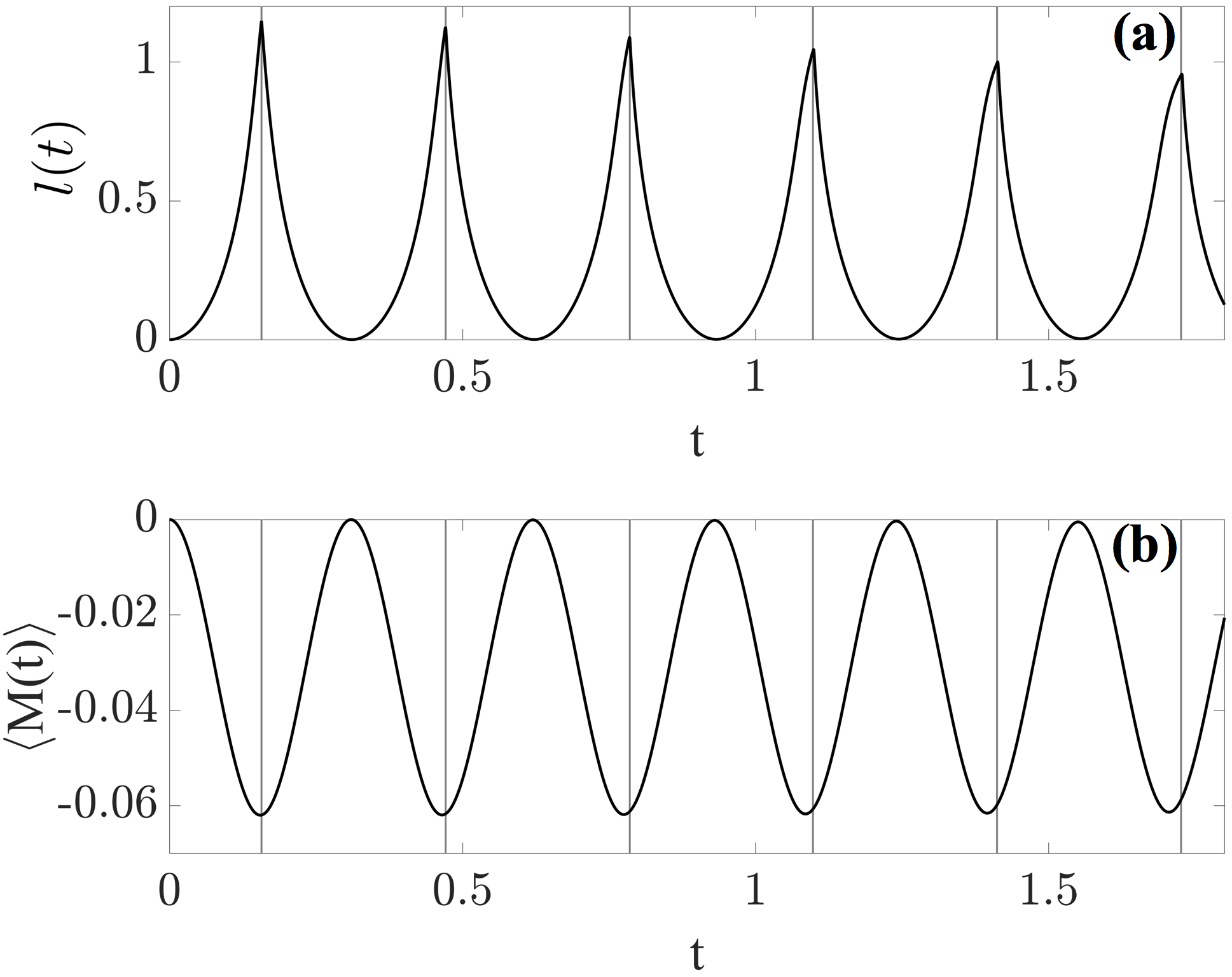}
\caption{(a) $l(t)$ and (b) $\langle M(t) \rangle$ vs. $t$ for $N=100$ and a quench from superfluid ($V_i=0$) to deep in the insulating phase ($V_f=10$).
}
\label{Figure2}
\end{figure}

In order to distinguish between the superfluid and the insulating phase, we choose the experimentally accessible parity operator which is an observable that characterizes charge density wave order
\begin{equation}
M=\frac{1}{N}\sum_i(-1)^i b^{\dagger}_i b_i.
\end{equation}
In the fermionic representation this is given by
\begin{eqnarray}
M=\frac{1}{N}\sum_{|k|<\pi/2}\hat{\Psi}^{\dagger}(k)\;\hat{\sigma}_x\;\hat{\Psi}(k).
\end{eqnarray} 
The calculation of $\langle M\rangle$ 
gives
\begin{eqnarray}
&&\langle M(t)\rangle= -\frac{1}{N}\sum_{|k|<\pi/2}\sin(2\theta_k(V_f))\cos(2\Delta\theta_k) \nonumber \\
&&+ \frac{1}{N}\sum_{|k|<\pi/2}\cos(2\theta_k(V_f))\sin(2\Delta\theta_k)\cos(2\epsilon_k(V_f)t).   
\end{eqnarray} 
Plotting this function in general in a situation where singularities in the rate function are present
shows that while both quantities oscillate the time scales are typically very different (see Fig(\ref{Figure3})). 
There is however one instance in which the two quantities 
appear to have a correlated behavior (see Fig.(\ref{Figure2})), i.e. for quenches from $V_i\leq0$ to a large $V_f>0$ ($V_f/J\gg1$). In this case
the parity operator oscillates between a positive and negative value periodically and each time a minimum is 
attained a cusp singularity is observed. This result is however simple to understand: the  period of the oscillations of 
$\langle M(t)\rangle$ is $T_R=\pi/\sqrt{V_f^2+(2J)^2}$ (restoring the tunneling strength $J$), while that of the singularities is 
$T_R=\pi/\sqrt{V_f(V_f-V_i)}$. The two are clearly equal if $V_f\gg V_i, J$ in which case the dispersion is effectively flat and all 
$k$-modes oscillate with the same frequency. Therefore only in this case the orthogonality appears to be tied to oscillations of the
order parameter.
One might be tempted to argue that this is just the wrong operator to detect orthogonality. If however a different operator is used, such as for example the kinetic energy operator,
\be
K=\frac{1}{N}\sum_j\;b^{\dagger}_jb_{j+1}+h.c.
\ee
it is easy to show that after a quench
\be
K=&&-\frac{1}{N}\sum_{|k|<\pi/2} (2\cos(k))\Big[ \cos(2\theta_k)\cos(2\Delta\theta_k) \nonumber \\
&&+\sin(\theta_k)\sin(\Delta\theta_k)\cos(2\epsilon_kt) \Big],
\ee
which produces results similar to the ones presented above.

\section{Discussion}
In this paper we have undertaken an extensive study of dynamical criticality in systems which contain a superfluid-Mott insulator transition in equilibrium. Furthermore, to our knowledge, this is the first numerical study of this type in a continuum model. We have found that although non-analyticities are present for certain quenches which signal temporal orthogonality this is only manifested in experimentally relevant observables for deep quenches. We studied numerically the dynamics of both the rate function and the momentum distribution following a quench in the Tonks-Girardeau gas across the pinning transition. In the discrete case we provided analytic calculations for the rate function and the dynamics of the parity operator. As known from state discrimination protocols in quantum information, it is an extremely difficult task to uncover global orthogonality from local measurements on pure states \cite{walgate2000local} and in the case of mixed states it is generically impossible \cite{bennett1999quantum}, 
so we are lead to conjecture that in general it is not possible to detect orthogonality in the dynamics of the many-body state and hence non-analyticities in the rate functions by observing the dynamics of local observables alone. Nevertheless, we stress that one could still hope to detect such points through non-trivial order parameters \cite{budich2016dynamical} or perhaps even by extending ancilla based interferometry schemes which have been proposed \cite{goold2011orthogonality,knap2012time,sindona2013orthogonality, sindona2014statistics} and experimentally implemented in local quenches in Fermi gases \cite{cetina2016ultrafast}. In addition, studying the dynamics of the rate function and these experimentally relevant observables for quenches in critical models is interesting in its own right and we hope it will inspire further experiments in this direction.

{\bf Acknowledgements-} This work was supported by the Okinawa Institute of Science and Technology Graduate University. AS and JG would like to thank M. Heyl for discussions. JG would like to thank M. Collura for discussions.  JG in AS se zahvaljujeta ljubljenim kra\v{s}kim gri\v{c}em in kolovozom, na katerih sta analizirala ve\v{c}ino znanstvenih rezultatov tega rokopisa v pogojih pomanjkanja kisika.
\bibliography{biblograph}

\end{document}